\documentclass[pre,showpacs,twocolumn]{revtex4}
\usepackage{graphicx}

\begin{document}

\title{Fluctuation theorem applied to the Nos\'e-Hoover thermostated
  Lorentz gas}
\author{Thomas Gilbert}
\email[]{thomas.gilbert@ulb.ac.be}
\affiliation{Center for Nonlinear Phenomena and
  Complex Systems,
  Universit\'e Libre de Bruxelles, Code Postal 231, Campus Plaine, B-1050
  Brussels, Belgium}
\date{\today}

\begin{abstract}
We present numerical evidence supporting the validity of the
Gallavotti-Cohen Fluctuation Theorem applied to the driven Lorentz gas with
Nos\'e-Hoover thermostating. It is moreover argued that the asymptotic form
of the fluctuation formula is independent of the amplitude of the driving
force, in the limit where it is small.
\end{abstract}

\pacs{05.70.Ln, 05.45.-a}

\maketitle 

Over the last decade, different versions of fluctuation formulas have
been the focus of a number of publications in the field of non-equilibrium
statistical physics. In particular, dissipative deterministic dynamical
systems with time-reversal symmetry have attracted some attention as potential 
candidates to model externally driven systems with a thermostating mechanism
\cite{ECM93,ES94,GC95}. Two distinct results have been
proposed, which, in the context of iso-kinetic thermostats, both
characterize the fluctuations of entropy production. One, due to Evans and
Searles \cite{ES94}, is usually referred to as \textit{transient
  fluctuation theorem}, while the other, due to Gallavotti and Cohen
\cite{GC95}, is simply known as the \textit{fluctuation theorem}. The
former addresses the fluctuations of the work done by the external forcing
on the system, and the latter the fluctuations of the phase space
contraction rate of non-equilibrium stationary states. It has been
rigorously proved in the context of Anosov systems \cite{GC95}. 

To be definite, consider the externally driven periodic Lorentz gas with
Gaussian thermostating \cite{MH87}. The trajectory of a particle in between
elastic collisions is described by the equation $\dot\mathbf{p} =
\mathbf{E} - \alpha \mathbf{p}$, where $\alpha =
\mathbf{E}\cdot\mathbf{p}/p^2$ is a reversible damping mechanism that acts
so as to keep the kinetic energy constant. $\alpha$, the phase space
contraction rate for this system is, as seen from its
expression, equal to the work done on the particle divided by the constant
temperature. Thus the work done on the particle is exactly compensated by
the heat dissipation. In other words, work and heat dissipation statistics
are identical for this system. 

Dolowschi\'ak and Kov\'acs \cite{DK05} recently made the observation that
work and phase space contraction rate fluctuations behave very differently
for the 
externally driven Lorentz gas with Nos\'e-Hoover thermostating. On the one
hand, the work fluctuations, whether large or small, obey the Evans-Searles
formula, in agreement with similar observations made for other systems 
\cite{ES00,ESR05}. On the other hand, the authors observed that the phase
space contraction rate fluctuations rapidly saturate. Moreover no
observation of a linear regime of fluctuations in a limited range was
reported. 

The Nos\'e-Hoover thermostated Lorentz gas on a periodic lattice has
phase space coordinates $\Gamma = (\mathbf{q}, \mathbf{p}, \zeta)$. Here
$\mathbf{q}$ and $\mathbf{p}$ respectively denote 
the position and momentum of the particle, and $\zeta$ is the variable
associated to the thermal reservoir.  Between two elastic collisions, the
dynamics is specified by the equations
\begin{equation}
\begin{array}{lcl}
\vspace{2mm}
\dot{\mathbf{q}} &=& \mathbf{p},\\
\vspace{2mm}
\dot{\mathbf{p}} &=& \mathbf{E} - \zeta \mathbf{p},\\
\dot\zeta &=& \tau_\mathrm{resp}^{-2}[p^2/(2T)-1].
\end{array}
\end{equation}
Here $\mathbf{E}$ denotes the external field,
$\tau_\mathrm{resp}$ the relaxation time of the thermostat and $T$ the
temperature \footnote{The role of $T$ as temperature is consistent as long
  as the external field is non zero. For zero field, the system has a
  limit-cycle. It is therefore not ergodic and the notion of temperature is
  ill-defined.}. 

An essential difference between the Gaussian and Nos\'e-Hoover thermostated
Lorentz gases is that the phase space of the former is compact. The
fluctuation theorem \cite{GC95} can thus be applied to the Gaussian
thermostated Lorentz gas as though it were Anosov \footnote{The Lorentz gas
or Gaussian thermostated driven Lorentz gas is strictly speaking not Anosov
because of the instantaneous collisions with the scatterers.}. For the
Nos\'e-Hoover thermostated Lorentz gas, the situation is different in that
the phase space contraction rate fluctuations are unbounded. In that case,
for large  fluctuations, one expects a much larger probability of positive
phase space contraction rate fluctuations.

Nevertheless, an appropriate modification of the fluctuation theorem was
given in \cite{BGGZ05}. Thus consider a time-reversible dissipative system
with average phase space contraction rate $\langle \sigma(\Gamma) \rangle =
\sigma_+ >0$ and assume this system verifies the chaotic hypothesis. In a
language similar to that used in \cite{DK05}, the statement of the
fluctuation theorem for the dimensionless contraction amplitude $p$ is that 
there exists a finite number $1\le p^* <\infty$, so that
\begin{equation}
\lim_{\tau\to\infty}\frac{1}{\tau}\log\frac{P_\tau(p)}{P_\tau(-p)} =
p  \sigma_+, \quad |p| < p^*,
\label{FT}
\end{equation}
where $P_\tau(p)$ denotes the probability of observing, over a time interval
$\tau$, a fluctuation of the phase space contraction rate 
$\langle \sigma \rangle_\tau = p\sigma_+$. In particular, thinking about
the external driving parameter in the Lorentz gas, the above result should
be independent of its amplitude ($|\mathbf{E}|>0$). 

We argue the driven periodic Lorentz gas with Nos\'e-Hoover thermostating
considered in \cite{DK05} verifies the fluctuation relation
Eq. (\ref{FT}); although \cite{DK05} correctly pointed out the phase space 
contraction rate fluctuations are bounded for large fluctuations,
their measurements do not point to the violation of Eq.~(\ref{FT}).
Here we present evidence that the Lorentz gas with Nos\'e-Hoover
thermostating has fluctuations of the phase space contraction rate which
are entirely consistent with the statement above, showing both saturation
for large fluctuations and linear behavior for small fluctuations.

As noted in \cite{BGGZ05}, the phase space contraction rate for this system
has the form
\begin{equation}
\sigma(\Gamma) = \sigma_0(\Gamma) + \frac{1}{T} \frac{d}{dt}H,
\end{equation}
where $\sigma_0(\Gamma) = \mathbf{E}\cdot\mathbf{p}/T$ is the quantity
relevant to the Evans-Searles fluctuation formula. Here $H = p^2/2 + T
\tau_\mathrm{resp}^2 \zeta^2$. Following the discussion in \cite{BGGZ05},
one can derive the distribution of $\sigma$ in terms of that of
$\sigma_0$. This was first done in \cite{vZC03} in the framework of a
Brownian particle dragged through water by a moving potential. In that
case, $\sigma_0$ has Gaussian fluctuations and one can derive the
assymptotic form of the RHS of Eq.~(\ref{FT})~:
\begin{equation}
\lim_{\tau\to\infty}\frac{1}{\tau}\log\frac{P_\tau(p)}{P_\tau(-p)} =
f(p)\sigma_+,
\label{FTGauss}
\end{equation}
where
\begin{equation}
f(p) = \left\{
\begin{array}{l@{\quad}l}
p , &0\le p < 1,\\
p - (p-1)^2/4, & 1\le p <3,\\
2, & p\ge 3.
\end{array}
\right.
\label{FTasymp}
\end{equation}
For negative $p$, $f(p)$ is odd, $f(-p) = - f(p)$.

\begin{figure}[htb]
\centering
\includegraphics[width=8cm]{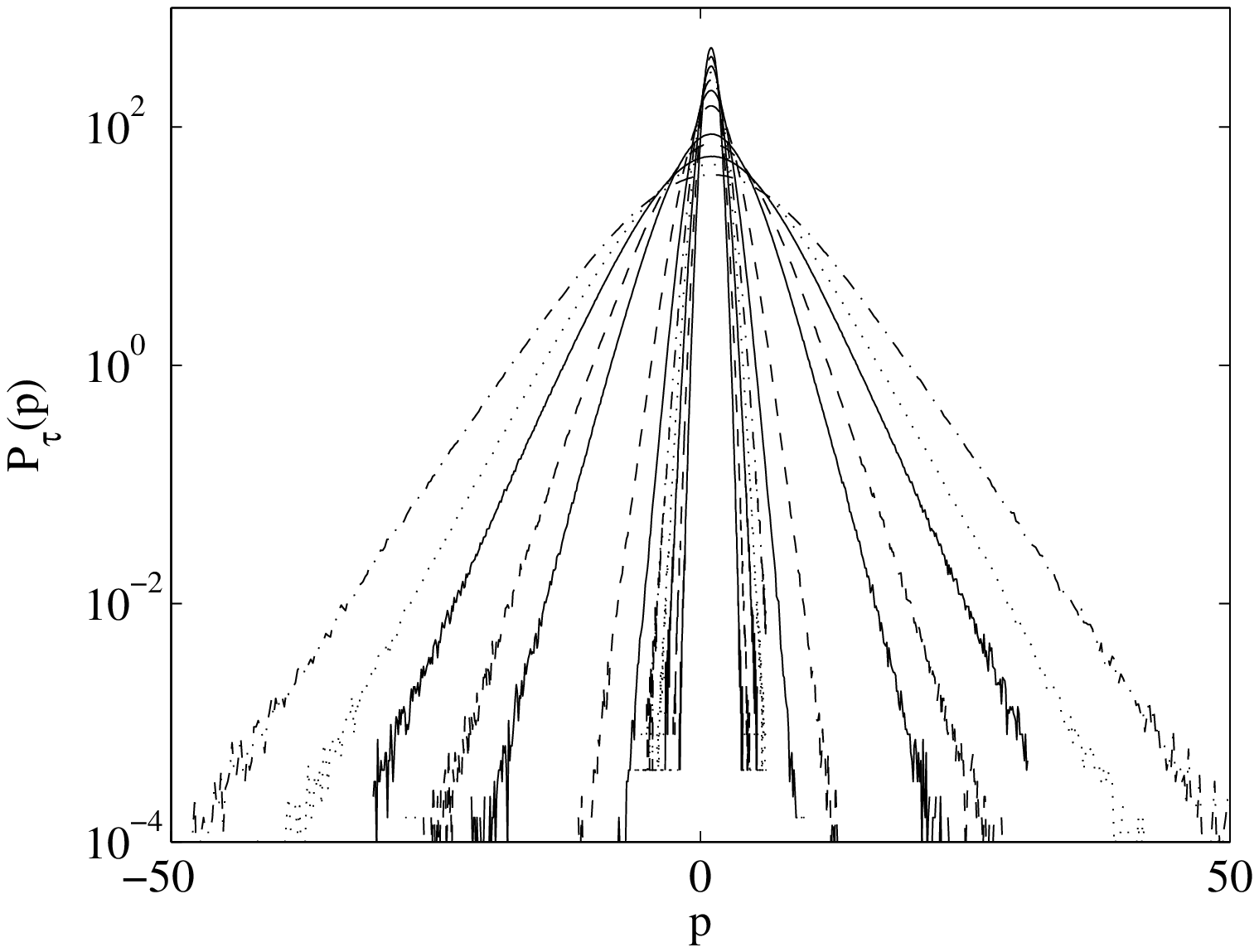}
\includegraphics[width=8cm]{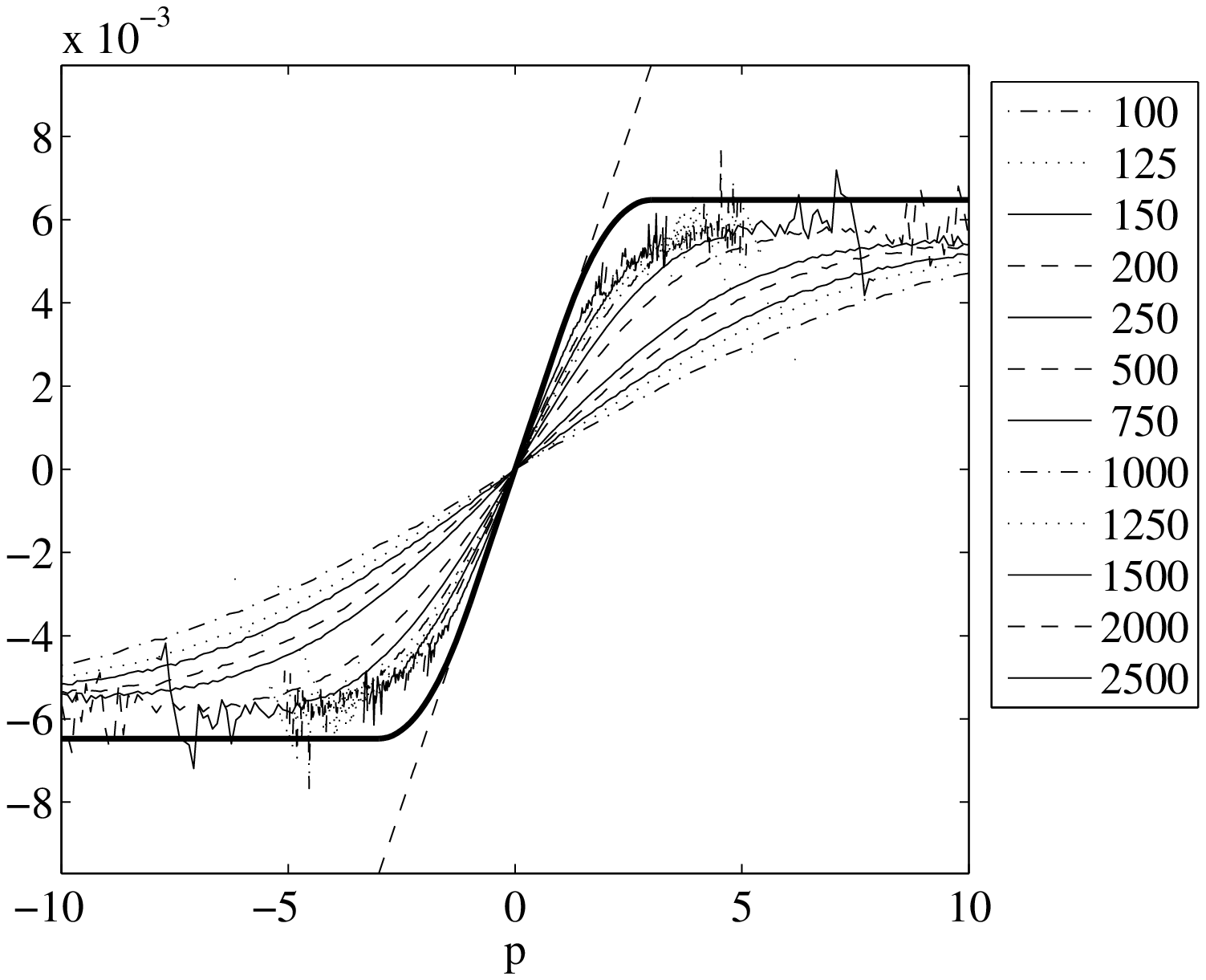}
\caption{Probability density function (above) and verification of
  Eq. (\ref{FT}) (below) for different time averages, as indicated in the
  legend. The parameters are set to $\mathbf{E} = (0.1,0)$, $T=1/2$
  and $\tau_\mathrm{resp} = 1$. The average phase space contraction rate
  $\langle 2\zeta \rangle = 3.25\times 10^{-3}$ was measured over a total
  of $\approx 3\times10^8$ collisions. The thick solid curve corresponds to
  the prediction in Eqs.~(\ref{FTGauss})-(\ref{FTasymp}).}
\label{fig.FT.1}
\end{figure}

\begin{figure}[tbh]
\centering
\includegraphics[width=8cm]{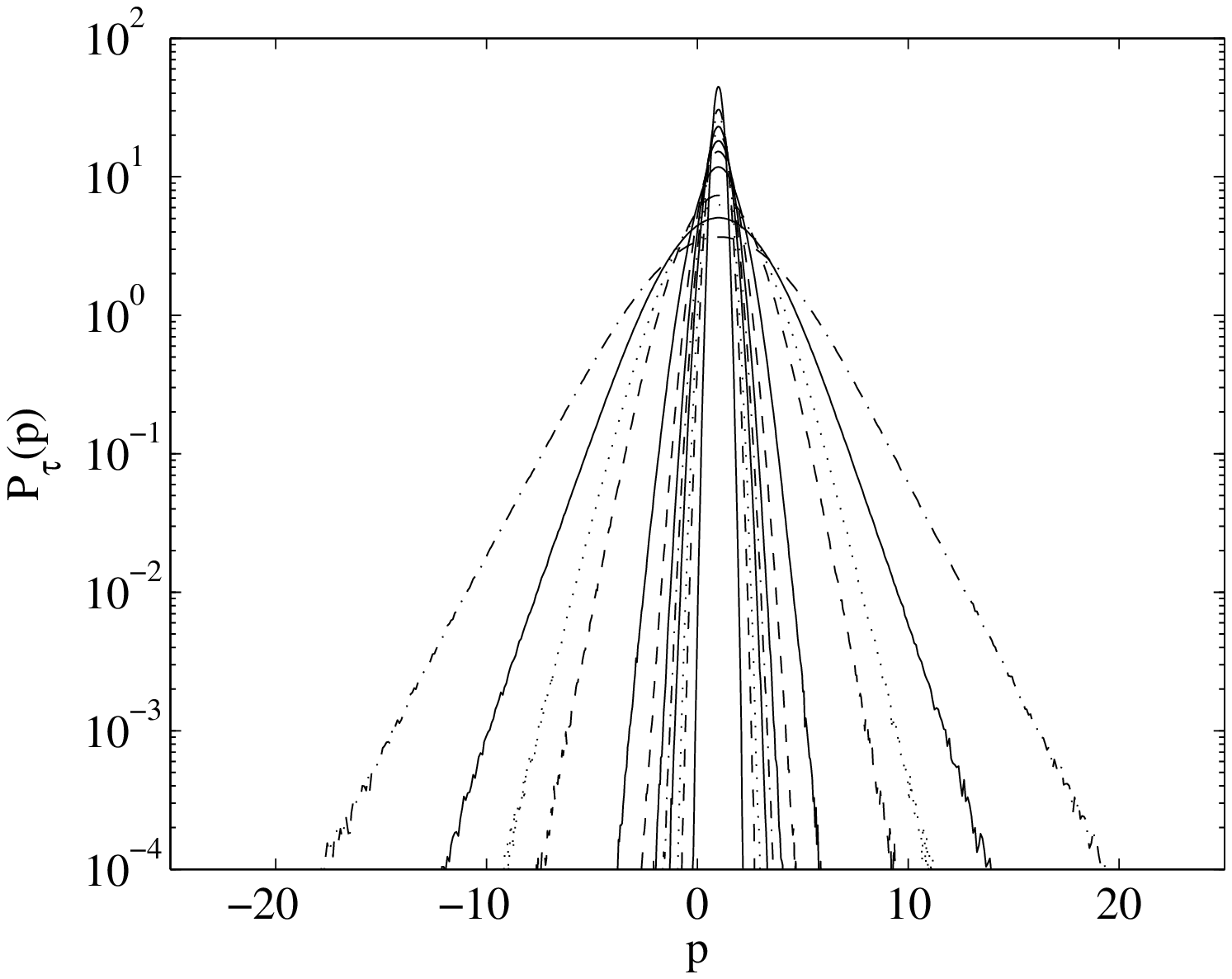}
\includegraphics[width=8cm]{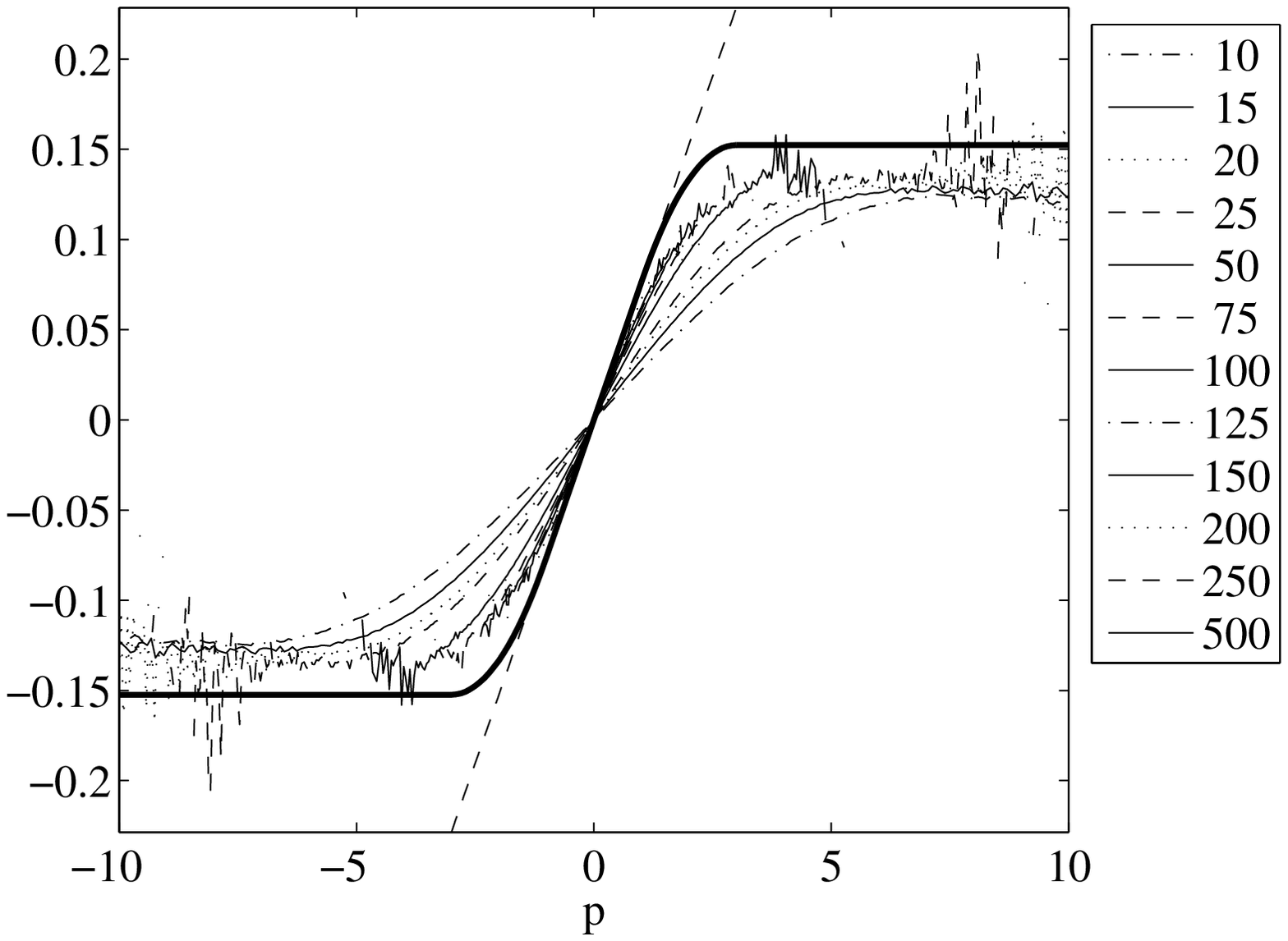}
\caption{Same as Fig. \ref{fig.FT.1} for $\mathbf{E} = (0.5,0)$. The
  average phase space contraction rate $\langle 2\zeta \rangle = 7.62\times
  10^{-2}$ was measured over a total of $\approx 3\times10^8$ collisions.} 
\label{fig.FT.5}
\end{figure}

According to \cite{DK05}, the fluctuations of the time averages of
$\sigma_0$ for the Nos\'e-Hoover Lorentz gas are Gaussian \footnote{
We could check this observation to very good accuracy.
For instance, the distributions of $\sigma_0$ corresponding to the
data presented in Figs. \ref{fig.FT.1}-\ref{fig.FT.5} have kurtosis (fourth
moment divided by second moment squared) equal to 3, the Gaussian
value, within one percent (and similarly for higher
moments 6 and 8). Likewise the skewness (third moment
squared by second moment to the cube) is zero within
one thousandth.}, which, given that the kinetic energy probability distribution
is canonical in the presence of the Nos\'e-Hoover thermostat, entitles
us to use the result of \cite{vZC03}, Eqs.~(\ref{FTGauss})-(\ref{FTasymp}).
We present in Figs. \ref{fig.FT.1}-\ref{fig.FT.5} the results of numerical
simulations on a hexagonal lattice with inter cell distance unity and disk
radius $0.44$ (consistent with the finite horizon condition). That is we
take the fundamental lattice translation vectors to be $(1,0)$ and
$(1/2,\sqrt{3}/2)$. Two different values of the external field are 
considered, both along the $x$-axis, $\mathbf{E} = (0.1,0)$
and $\mathbf{E} = (0.5,0)$. The numerical integration
was performed using an algorithm similar to that used in \cite{RKN00}.

\begin{figure}[tbh]
\centering
\includegraphics[width=8cm]{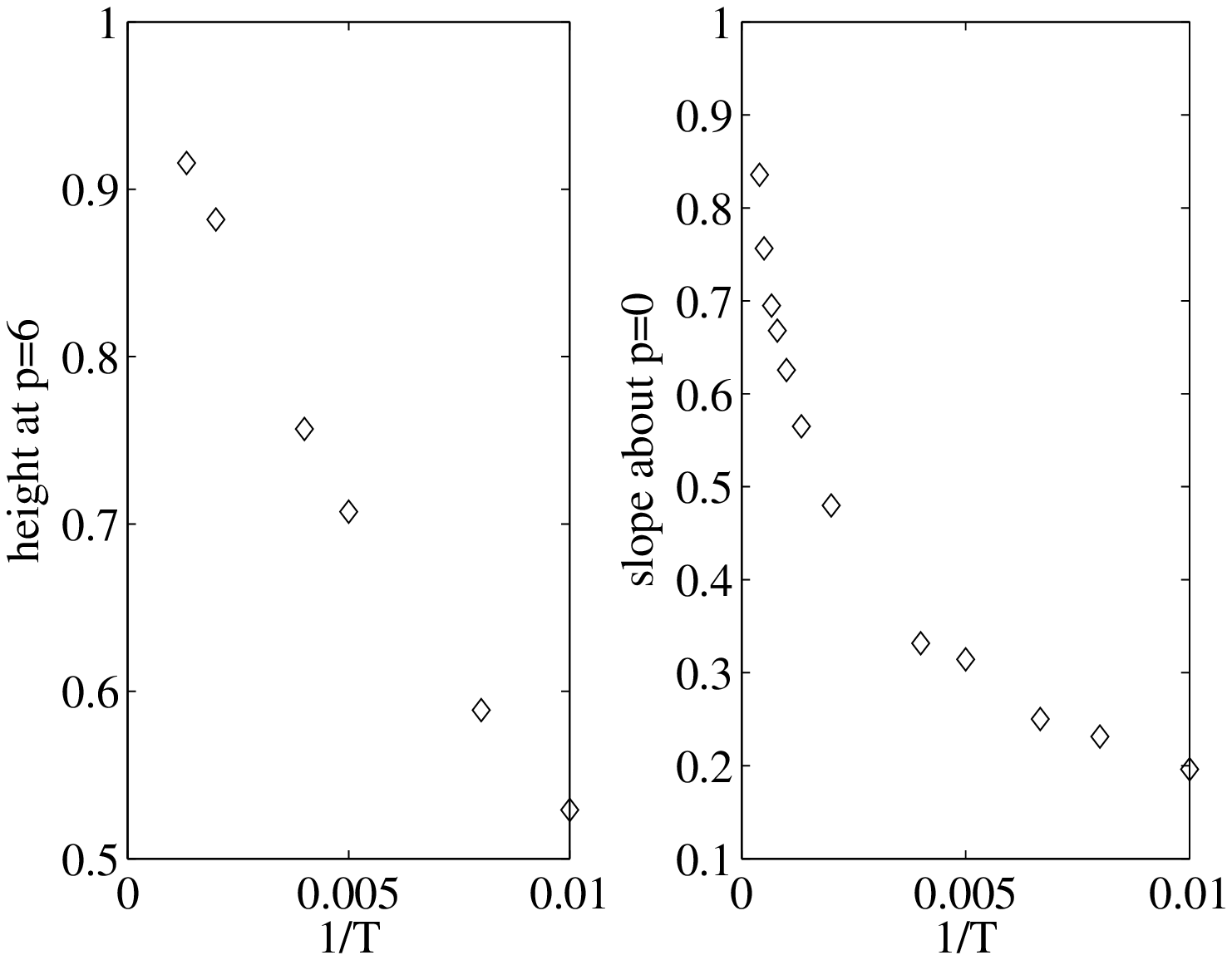}
\caption{Height of saturation measured at $p = 6$ (for times $T =
  100,\dots, 750$) (left), and slope measured about $p = 0$ vs.$1/T$ of the
  curves shown in the bottom-panel of Fig. \ref{fig.FT.1} for the external
  forcing $\mathbf{E} = (0.1, 0)$. The vertical coordinates are
  renormalized to the asymptotic values predicted by
  Eq. (\ref{FTasymp}). Both curves can be linearly extrapolated to 1 as
  $T\to\infty$. These data confirm that $\mathcal{O}(1/T)$ corrections
  affect the finite time measurements. }
\label{fig.FT.apx}
\end{figure}

As seen from the figures, the numerical data is entirely consistent with
Eq. (\ref{FT}). Indeed, as times become large, the small fluctuation
amplitudes have a linear slope which approaches asymptotically
the value given by the RHS of Eq. (\ref{FT}). The data
are compared to the prediction in Eqs. (\ref{FTGauss})-(\ref{FTasymp}) and
show a rather good agreement, albeit $f(p) = 2$ seems to slightly
overestimate the saturation level. As illustrated on the
left-panel of Fig. \ref{fig.FT.apx}, this apparent discrepancy is only a
finite-time correction, expected to be $\mathcal{O}(1/T )$
\cite{GZG05}. Likewise the right-panel of Fig. \ref{fig.FT.apx} shows that
the slope in the linear regime of fluctuations approaches the predicted
asymptotic value as average lengths
increase. Figs. \ref{fig.FT.1}-\ref{fig.FT.5} 
moreover show the linear regime of fluctuations persists
irrespective of the strength of the driving field. That is
to say, the field strength is relevant to the asymptotic
regime only through $\sigma_+$ which is quadratic in the field
strength, consistent with Eqs. (\ref{FTGauss})-(\ref{FTasymp}). Similar results
were obtained for field values as low as  $\mathbf{E} = (0.05,0)$.
The main difficulty being that the length of the time averages
in Eq. (\ref{FT}) needs to be increased significantly in
order for the slope to converge to the asymptotic value
predicted by Eq. (\ref{FT}).

In summary, the driven periodic Lorentz gas with Nos\'e-Hoover
thermostating provides a simple example of a fully deterministic model
whose work and phase space contraction rate fluctuations obey relations
similar to those 
found in \cite{vZC03} for the work and heat fluctuations of a Brownian
particle in a potential well. Similar results 
were obtained in \cite{GC05} for the non-equilibrium fluctuations of a RC
circuit. This suggests an identification between phase space contraction
rate and heat dissipation in this framework. The numerical evidence we
presented is entirely consistent with the phase space contraction rate
fluctuation theorem as stated in \cite{BGGZ05}. The sharp constrast between
this observation and the conclusion drawn by \cite{DK05} (similar claims can
be found in \cite{ESR05}) that the fluctuation theorem does not hold if the
driving is small can be attributed to two reasons; (i) these authors did
not consider properly the saturation of the fluctuation at larger $p >
p^*$, and (ii) they did not run the simulation long enough to observe the
small but clear linear region at $p < p^*$. The small deviations between
our data and the form predicted by Eqs. (\ref{FTGauss})-(\ref{FTasymp})
should be attributed to finite-time corrections. We hope to further report
on this issue in a future publication.

\begin{acknowledgments}
The author wishes to thank M. Dolowschi\'ak, Z. Kov\'acs, G. Nicolis,
G. Gallavotti, and R. van Zon  for helpful comments, as well as F. Zamponi
who pointed out the 
relevance of Eqs.~(\ref{FTGauss})-(\ref{FTasymp}). The author would also
like to thank  P. Gaspard and J.~R. Dorfman for their continuing 
support. The author is charg\'e de recherches with the F.~N.~R.~S. (Belgium). 
\end{acknowledgments}

\end{document}